\documentstyle[12pt]{article}
\begin{document}
\pagestyle{empty}
\begin{flushright}
{ \bf SNB/HEP/99-03-01}\\
{\bf March 1999}
\end{flushright}
\vskip .7cm
\begin{center}

{\bf {TOPOLOGICAL ASPECTS IN NON-ABELIAN GAUGE THEORY}}

\vskip 2cm

{\bf R. P. Malik} 
\footnote{On leave of absence from Bogoliubov Laboratory of
     Theoretical Physics, JINR-Dubna, Moscow, Russia.\\
e-mail addresses: malik@boson.bose.res.in AND  malik@thsun1.jinr.dubna.su}\\
{\it S. N. Bose National Centre for Basic Sciences,}\\
{\it Block-JD, Sector III, Salt Lake, Calcutta- 700 091, India.}\\

\vskip .5cm

{\bf ABSTRACT}

\end{center}

We discuss the BRST cohomology  and exhibit a connection
between the Hodge decomposition theorem and the topological properties
of a two dimensional free non-Abelian gauge theory having
no interaction with matter fields. The topological nature of this
theory is encoded in the vanishing of the Laplacian operator when
equations of motion are exploited.
We obtain two sets of topological invariants with respect to BRST
and co-BRST charges 
on the two dimensional manifold and show that the
Lagrangian density of the theory can be expressed as the sum of
terms that are BRST- and co-BRST invariants.\\

\baselineskip=16pt

\vskip 1cm
\newpage

\pagestyle{headings}

\noindent 
{\bf 1 Introduction}\\

\noindent
The local gauge theories, endowed with the first class constraints [1,2],
play a key role in the understanding of the basic interactions of 
nature (except quantum gravity). For the Becchi-Rouet-Stora-Tyutin (BRST) 
quantization of such a class of theories, the local gauge symmetry
transformations of the classical theories are traded with the quantum gauge
BRST symmetry transformations which are 
generated by the conserved ($ \dot Q_{B} = 0$) and nilpotent
($ Q_{B}^2 = 0$) BRST charge $Q_{B}$ [3-5].  In particular, the importance 
of the BRST formalism comes to its full glory in the context of the 
covariant canonical quantization of the non-Abelian gauge theory
where unitarity and gauge invariance both are respected together. 
The presence of the first class 
constraints of the original theories is found to be 
encoded in the subsidiary
condition $ Q_{B} |phys > = 0$ which implies that the physical states
are annihilated by these constraints.
The above two properties, i.e., the physical state
condition $Q_{B} |phys> = 0 $ and the nilpotency of the BRST charge
$Q_{B}^2 = 0$, are the two key requirements to define the cohomological
aspects of the BRST formalism. The
inclusion of the BRST symmetry in the Batalin-Vilkovisky formalism
(see, e.g., [6,7]), the  discussion of  the second class constraints in
its framework [8], its indispensible use in the topological
field theories [9-11] and string theories, etc., have enriched
the physical and  mathematical aspects of the BRST formalism to
a fairly high degree of sophistication.

One of the most celebrated theorems in the mathematical
aspects of the de Rham cohomology is
the Hodge decomposition theorem defined on a compact manifold [12-14].
This theorem  states that any arbitrary form can be 
written as the sum of a harmonic form, an exact form and a 
co-exact form. In principle, the cohomology can be defined w.r.t.
the exterior derivative $d (d^2= 0)$ and/or w.r.t. the dual extrerior
derivative $\delta (\delta^2 = 0)$ where $\delta = \pm * d *$ is the
Hodge dual of $d$.  The operation of $d$ on any arbitrary form 
increases the degree of the form by one whereas the operation of
$\delta$ reduces it by one. In the cohomological description
of BRST formalism, the conserved and nilpotent BRST charge (which
generates a nilpotent quantum gauge symmetry) is identified with the
exterior derivative $d$ in any arbitrary 
dimensions of spacetime. It would be, therefore, an 
interesting endeavour to obtain the analogues of 
$\delta$ and the Laplacian $\Delta = (d + \delta)^2
= d \delta + \delta d $ in the language
of the nilpotent (for $\delta$), local, covariant and 
continuous symmetry properties of a given Lagrangian density. 
Some attempts [15] have been made
towards this goal in four dimensions of spacetime but the
symmetry transformations turn out to be nonlocal and noncovariant. 
In the covariant formulation, the symmetry transformations turn out to
be even non-nilpotent and they become nilpotent only under
certain specific restrictions [16].

The central theme of the present paper is to express the Hodge
decomposition theorem in terms of the local and conserved charges
corresponding to the analogues of $d$, $\delta$ and $\Delta$ of
differential geometry and establish a connection with the topological
nature of the two ($ 1 + 1$) dimensional non-Abelian gauge theory
having no interaction with matter fields. We generalise our works
for the free Abelian $U(1)$ gauge theory in two dimensions (2D) 
[17,18] and
show that the Laplacian operator for the free non-Abelain gauge theory
(without any interaction with matter fields) too, goes to zero when 
equations of motion are exploited. This happens due to the fact that
both the physical degrees of freedom of the non-Abelian gauge boson
are gauged away by the presence of the nilpotent BRST- and 
co(dual)- BRST symmetries in the theory. Thus,  theory becomes topological
in nature (see, e.g., Ref. [11]). Mathematically, the Lagrangian 
density of the theory can be expressed as the sum of  
BRST- and co-BRST invariant parts and, therefore, it bears
an outlook similar to the Witten type theories [10].
The topological nature of this theory is confirmed by the
existence of two sets of topological invariants on the 2D compact
manifold. In Sec. 2, we set up the notations and give 
a concise description of the BRST formalism for the free 
D-dimensional non-Abelian gauge theory.
This is followed by the derivation of the (anti)dual BRST 
symmetries and corresponding charges in Sec. 3. In Sec. 4,
we obtain the symmetries that are generated by the analogue of
the Laplacian(Casimir) operator and derive the full BRST algebra. 
Sec. 5 is devoted to the discussion of Hodge decomposition 
theorem and the derivation
of the topological invariants. Finally, we make some 
concluding remarks and point out some directions that
can be pursued in the future.\\

\noindent
{\bf 2 Preliminary: BRST symmetries}\\

\noindent
Let us begin with the BRST invariant Lagrangian density (${\cal L}_{B}$) 
for the D-dimensional free non-Abelian gauge theory (having no interaction
with matter fields)
in the Feynman gauge
$$
\begin{array}{lcl}
{\cal L}_{B} = - \frac{1}{4} F^{\mu\nu a} F^{a}_{\mu\nu}
+ B^a (\partial \cdot A)^a + \frac{1}{2} B^a B^a
- i \partial_{\mu}\bar C^a D^{\mu} C^a,
\end{array} \eqno(2.1)
$$
where
$F^{a}_{\mu\nu} = \partial_{\mu} A^{a}_{\nu} 
- \partial_{\nu} A^{a}_{\mu} + g f^{abc} A^b_{\mu} A^c_{\nu}$ 
is the field strength tensor obtained
from the group valued gauge connection $A^a_{\mu}$,
$B^a$ is the group valued auxiliary field,
$(\bar C^a) C^a$ are the (anti)ghost fields ($(\bar C^{a})^2
= (C^a)^2 = 0$) and the covariant derivative is defined as:
$ D_{\mu} C^a = \partial_{\mu} C^a
+ g f^{abc} A^b_{\mu} C^c $, the D-dimensional
flat Minkowski spacetime indices are $\mu,\nu.....
= 0, 1, 2,...D-1$, the group indices 
 $ a,b, c,....= 1, 2, 3,...$ correspond to the compact Lie gauge group,
 $g$ is the coupling constant and 
structure constants $f^{abc}$ are chosen to be totally
antisymmetric in $a,b,c$ (see, e.g., [19]). 
The above Lagrangian density respects ($ \delta_{B} 
{\cal L}_{B} = \eta \partial_{\mu} [ B^a D^\mu C^a ] $) off-shell
nilpotent $(\delta_{B}^2 = 0)$ BRST symmetry transformations 
$$
\begin{array}{lcl}
\delta_{B} A^a_{\mu} &=& \eta D_{\mu} C^a, \;\;\;\qquad \;\;\;
\delta_{B} C^a = - \frac{\eta g}{2} f^{abc} C^b C^c,\;\;\;
\qquad \;\;\;\;\delta_{B} B^a = 0, \nonumber\\
\delta_{B} F^a_{\mu\nu} &=& \eta g f^{abc} F^b_{\mu\nu} C^c,\;\quad\;
\delta_{B} \bar C^a = + i \eta B^a, \;\quad\;
\delta_{B} (\partial \cdot A)^a = \eta \partial_{\mu} D^{\mu} C^a,
\end{array} \eqno(2.2)
$$
where $\eta$ is an anticommuting $( \eta C^a = - C^a \eta, \eta \bar C^a
= - \bar C^a \eta)$ spacetime independent transformation parameter.
The on-shell ($ \partial_{\mu} D^{\mu} C^a = 0$) nilpotent 
($\delta_{b}^2 = 0$) BRST transformations
$$
\begin{array}{lcl}
\delta_{b} A^a_{\mu} &=& \eta \;D_{\mu} \;C^a, \qquad
\;\delta_{b} F^a_{\mu\nu} = \eta g f^{abc}
F^b_{\mu\nu} C^c, \;\qquad \;\; 
\delta_{b} \bar C^a = - i \eta (\partial \cdot A)^a, \nonumber\\
\delta_{b} C^a &=&  - \frac{\eta g}{2} f^{abc} C^b C^c,
\qquad \;\delta_{b}(\partial \cdot  A)^a
= \eta \;\partial_{\mu} D^{\mu} C^a,
\qquad\; \delta_{b} ( D_{\mu} C^a ) = 0,  
\end{array} \eqno(2.3)
$$
can be derived from (2.2) by using the equation of motion
$ B^a = - (\partial \cdot A)^a $ and they 
leave the following Lagrangian density:
$$
\begin{array}{lcl}
{\cal L}_{b} = - \frac{1}{4} F^{\mu\nu a} F^a_{\mu\nu} 
- \frac{1}{2} (\partial \cdot A)^a (\partial \cdot A)^a 
- i \partial_{\mu} \bar C^a D^{\mu} \;C^a,
\end{array} \eqno(2.4)
$$
quasi-invariant because
$ \delta_{b} {\cal L}_{b} = - \eta
\partial_{\mu} [ (\partial \cdot A)^a D^\mu C^a ]$. These symmetries
lead to the following expression for the 
conserved and nilpotent BRST charge ($ Q_{(B,b)}$) (see, e.g., [3-5]):
$$
\begin{array}{lcl}
Q_{(B,b)} &=& 
 {\displaystyle \int d^{D-1} x}\; 
\bigl [\;B^a D_{0} C^a - \dot B^a C^a + \frac{1}{2} i g
f^{abc}\; \dot {\bar C^a}\; C^b\; C^c \;\bigr ], \nonumber\\
 &\equiv& {\displaystyle \int d^{D-1} x}\; 
\bigl [\; \partial_{0} (\partial \cdot A)^a C^a -
(\partial \cdot A)^a D_{0} C^a 
+ \frac{1}{2} i g f^{abc}\; \dot {\bar C^a}\; C^b\; C^c 
\; \bigr ].
\end{array} \eqno(2.5)
$$
The continuous global symmetry invariance
of the total action under the transformations: 
$ C^a \rightarrow e^{-\lambda} C^a,
\bar C^a \rightarrow e^{\lambda} \bar C^a, A^a_{\mu} 
\rightarrow A^a_{\mu}, B^a \rightarrow B^a $, 
(where $\lambda$ is a global parameter), leads
to the derivation of the conserved ghost charge $(Q_{g})$
$$
\begin{array}{lcl}
Q_{g} = - i {\displaystyle \int d^{(D-1)} x }\;
\bigl [\; C^a\; \dot {\bar C^a} + \bar C^a\; D_{0} C^a \;\bigr ].
\end{array}\eqno(2.6)
$$
The derivation of
the anti-BRST charge in non-Abelian gauge theory is more involved. 
In fact, one introduces
another auxiliary field $ \bar B^a $ in (2.1) for this purpose:
$$
\begin{array}{lcl}
{\cal L}_{\bar B} = - \frac{1}{4} F^{\mu\nu a} F^{a}_{\mu\nu}
+ B^a (\partial \cdot A)^a + \frac{1}{2} (B^a B^a + \bar B^a \bar B^a)
- i \partial_{\mu}\bar C^a D^{\mu} C^a,
\end{array}\eqno(2.7a)
$$
$$
\begin{array}{lcl}
{\cal L}_{\bar B} = - \frac{1}{4} F^{\mu\nu a} F^{a}_{\mu\nu}
- \bar B^a (\partial \cdot A)^a + \frac{1}{2} (B^a B^a 
+ \bar B^a \bar B^a)
- i D_{\mu}\bar C^a \partial^{\mu} C^a,
\end{array} \eqno(2.7b)
$$
where the auxiliary fields are restricted to satisfy [20]
$$
\begin{array}{lcl}
B^a + \bar B^a = i g \;f^{abc}\; C^b \;\bar C^c.
\end{array}\eqno(2.8)
$$
Under the BRST transformations, it is interesting to note that $\bar B^a$
transforms as: $ \delta_{B} \bar B^a = \eta g f^{abc} \bar B^b C^c $.
The following off-shell nilpotent ($\delta_{AB}^2 = 0$) 
anti-BRST ($\delta_{AB}$) transformations
$$
\begin{array}{lcl}
\delta_{AB} A^a_{\mu} &=& \eta D_{\mu} \bar C^a, \;\;\;\qquad \;\;\;\;
\delta_{AB} \bar C^a = - \frac{\eta g}{2} f^{abc} 
{\bar C}^b {\bar C}^c,\;\;\;
\qquad \;\;\;\delta_{AB} \bar B^a = 0, \nonumber\\
\delta_{AB} F^a_{\mu\nu} &=& \eta g f^{abc} 
F^b_{\mu\nu} \bar C^c,\;\quad\;
\delta_{AB}  C^a = + i \eta  \bar B^a, \;\;\quad\;\;
\delta_{AB} B^a = \eta g f^{abc} B^b \bar C^c,
\end{array} \eqno(2.9)
$$
leave the Lagrangian density (2.7b) quasi-invariant 
($ \delta_{AB} {\cal L}_{\bar B} = - \eta \partial_{\mu}
[\bar B^a D^{\mu} \bar C^a]$). The above transformations (2.9)
are generated by the nilpotent and conserved anti-BRST charge
$$
\begin{array}{lcl}
Q_{AB} &=& 
 {\displaystyle \int d^{D-1} x}\; 
\bigl [\;\dot {\bar B^a} \bar C^a
- \bar B^a D_{0} \bar C^a  - \frac{1}{2} i g
f^{abc} \dot { C^a} \bar C^b \bar C^c
\; \bigr ].
\end{array}\eqno(2.10)
$$
Together, the above three conserved charges satisfy:
$$
\begin{array}{lcl}
&& \{ Q_{B}, Q_{B} \}\;\;\; =\;\;\; \{ Q_{AB}, Q_{AB} \}\;\; \;= \;\;0, 
\nonumber\\
&& \{ Q_{B},   Q_{AB} \} = Q_{B} Q_{AB}\; +\; Q_{AB} Q_{B} = 0,\nonumber\\
&& i [  Q_{g}, Q_{B} ] = + Q_{B}, \quad 
i [  Q_{g}, Q_{AB} ] = - Q_{AB}, 
\end{array}\eqno(2.11)
$$
where the basic canonical (anti)commutators for the BRST 
invariant Lagrangian densities have been exploited. It can be seen
that the specific combinations of transformations: $(\delta_{B}
\delta_{AB} + \delta_{AB} \delta_{B})$ acting on any field generate
no transformation at all
(as $ \{ Q_{B}, Q_{AB} \} = 0 $). In particular, it can be checked that
$\{ \delta_{B}, \delta_{AB} \} A_{\mu}^a = 0 $ is obeyed if and only if
the restriction (2.8) is satisfield. For
the non-Abelian compact Lie algebra we have considered, the 
anticommutator of the BRST- and anti-BRST charges is zero 
\footnote{ In a recent work [21], it has been 
pointed out that the
cohomologically higher order BRST- and anti-BRST operators do not
anticommute and their anticommutator leads to the definition of
a cohomologically higher order Laplacian operator.}. 
Thus, the anti-BRST charge ($Q_{AB}$) is not the analogue of 
the dual exterior derivative ($\delta$).\\

\noindent
{\bf 3 Dual BRST symmetries}\\

\noindent
We consider here a two ($ 1 + 1$) dimensional non-Abelian gauge theory
and discuss dual BRST- and anti-dual BRST symmetries. 
The Lagrangian density (2.4) in 2D
\footnote{ We adopt here the notations in which the 2D flat Minkowski
metric is : $\eta_{\mu\nu} = ( +1, -1) $ and $ \Box = \eta^{\mu\nu}
\partial_{\mu} \partial_{\nu} = \partial_{0} \partial_{0} -
\partial_{1} \partial_{1},\; \dot \phi^a = \partial_{0} 
\phi^a, \;F^a_{01} = 
\partial_{0} A^a_{1} - \partial_{1} A^a_{0} + g f^{abc} A^b_{0} A^c_{1}
= E^a
= F^{10a},\; \varepsilon_{01} = \varepsilon^{10} = +1, 
\;(\partial \cdot A)^a = \partial_{0} A^a_{0} - \partial_{1} A^a_{1},
\;D_{\mu} \phi^a = \partial_{\mu} \phi^a + g f^{abc} A_{\mu}^b \phi^c,
\;D_{\mu} (\phi^a \psi^b) = D_{\mu} \phi^a \psi^b
+ \phi^a D_{\mu} \psi^b$. }
$$
\begin{array}{lcl}
{\cal L}_{b} =
\frac{1}{2} E^a E^a 
- \frac{1}{2} (\partial \cdot A)^a (\partial \cdot A)^a 
- i \partial_{\mu}
\bar C^a D^{\mu} C^a, 
\end{array} \eqno (3.1)
$$  
remains quasi-invariant ($ \delta_{d} {\cal L}_{b} = \eta
\partial_{\mu} [ E^a \partial^{\mu} \bar C^a ]$)
under the following on-shell ($ D_{\mu} \partial^{\mu} 
\bar C^a = 0$) nilpotent ($\delta_{d}^2 = 0$) symmetry transformations

$$
\begin{array}{lcl}
&&\delta_{d} A^a_{\mu} = - \eta \varepsilon_{\mu\nu} \partial^{\nu} \bar C^a,
\quad \;\delta_{d} \bar C^a = 0, \;\quad \;
  \delta_{d} C^a = - i \eta E^a,\;\quad \;\;
\delta_{d} (D_{\mu} \partial^\mu \bar C^a) = 0, \nonumber\\
&&\delta_{d} (\partial \cdot A)^a = 0, \qquad
\delta_{d} E^a = \eta D_{\mu} \partial^{\mu} \bar C^a,\qquad
\delta_{d} F^a_{\mu\nu}= \eta ( \varepsilon_{\mu\rho} D_{\nu}
- \varepsilon_{\nu\rho} D_{\mu} ) \partial^{\rho} \bar C^a. 
\end{array}\eqno(3.2)
$$
We christen this symmetry as the dual BRST symmetry by taking analogy
with the Abelian gauge theory where, like the above transformations,
it is the gauge-fixing
term $(\partial \cdot A)^a$ that remains invariant [17,18]. 
At this stage, it is essential to
pin-point some of the differences and similarities between the BRST- and
dual BRST symmetries in Abelian $U(1)$ gauge theory and the same
in the context of non-Abelian gauge theory. In the Abelian theory,
the gauge-fixing term $\delta A = (\partial
\cdot A) $ with $ \delta = \pm {*} d {*} $ is the Hodge dual of the two-form 
$ F = d A$ which is the electric field $E$ in 2D. This is not
the case, however, for the non-Abelian gauge theory because the
field strength tensor $F^a_{\mu\nu}$ contains a self-interacting 
term $ g f^{abc} A^b_{\mu}
A^c_{\nu} $ which is not present in the two-form $ F = d A $ of the
Abelian gauge theory. Under the BRST transformations 
in the Abelian gauge theory, it is the two-form $ F = d\; A$
that remains invariant $( \delta_{B} F_{\mu\nu} = 0)$
but for the non-Abelian gauge theory the field strength tensor transforms:
$ \delta_{B} F_{\mu\nu}^a = \eta g f^{abc} F_{\mu\nu}^b C^c $.
It is the total kinetic
energy term ($ - \frac{1}{4} F^{\mu\nu a}\;F^a_{\mu\nu}$), however,
that remains invariant in both kinds of 
gauge theories. Similarly, the dual BRST symmetry corresponds to a
symmetry in which the gauge-fixing term remains invariant. 
The analogue of the Lagrangian density (2.1) can
be written for the two-dimensional case by introducing one more
auxiliary field ${\cal B}^a$ as:
$$
\begin{array}{lcl}
{\cal L}_{\cal B} = {\cal B}^a E^a - \frac{1}{2} {\cal B}^a 
{\cal B}^a
+ B^a (\partial \cdot A)^a + \frac{1}{2} B^a B^a
- i \partial_{\mu}\bar C^a D^{\mu} C^a.
\end{array} \eqno(3.3)
$$
This Lagrangian density respects the off-shell nilpotent 
($ \delta_{D}^2 = 0$) dual BRST symmetry $\delta_{D}$
as well as the off-shell nilpotent ($\delta_{B}^2 = 0$)
BRST symmetry $\delta_{B}$. These symmetries, for
the above Lagrangian density,  are juxtaposed as 

$$
\begin{array}{lcl}
\delta_{D} A^a_{\mu} &=& - \eta 
\varepsilon_{\mu\nu} \partial^{\nu} \bar C^a,
\;\;\;\;\;\;\;\;\qquad \;\;\;\;\;\;\;\;\;\;
\delta_{B} A_{\mu}^a = \eta\; D_{\mu} C^a,
\nonumber\\
\delta_{D} \bar C^a &=& 0, \;\;\;\;\;\;\;
\;\;\;\;\;\;\;\;\;\;\;\qquad \;\;\;\;\;\;\;\;\;\;\;\;\;\;\;
\delta_{B} \bar C^a = i \eta B^a, \nonumber\\
 \delta_{D} C^a &=& - i \eta {\cal B}^a, 
\;\;\;\;\;\;\;\;\;\;\;\;\qquad\;\;\;\;\;\;\;\;\;\;\;\;\; 
\delta_{B} C^a = - \frac{\eta g}{2}
f^{abc} \;C^b\; C^c, \nonumber\\ 
\delta_{D} {\cal B}^a &=& 0, \;\;\;\;
\;\;\;\;\;\;\;\;\;\;\;\;\;\;\qquad \;\;\;\;\;\;\;\;\;\;\;\;\;\;\;
\delta_{B} {\cal B}^a = \eta g f^{abc} {\cal B}^b C^c, \nonumber\\
\delta_{D} (\partial \cdot A)^a &=& 0, \;\;\;\;\;\;\;\;\; 
\;\;\;\;\;\;\;\;\qquad \;\;\;\;\;\;\;\;\;\;\;\;\;\;\;\;
\delta_{B} (\partial \cdot A)^a = \eta \partial_{\mu} D^{\mu} C^a,
\nonumber\\
\delta_{D} B^a &=& 0, \;\;\;\;\;\;\;\;\;\;
\;\;\;\;\;\;\;\;\;\qquad\;\;\;\;\;\;\;\;\;\;\;\;\;\;
\delta_{B} B^a = 0,\nonumber\\
\delta_{D} E^a &=& \eta D_{\mu} \partial^{\mu} \bar C^a, 
\;\;\;\;\;\;\;\;\;\qquad\;\;\;\;\;\;\;\;
\;\;\; \delta_{B} E^a = \eta g f^{abc} \;E^b\;C^c, \nonumber\\
\delta_{D} F^a_{\mu\nu} &=& \eta (\varepsilon_{\mu\rho} D_{\nu}
- \varepsilon_{\nu\rho} D_{\mu}) \partial^{\rho} \bar C^a,\qquad
\delta_{B} F^a_{\mu\nu} = \eta g f^{abc} F^b_{\mu\nu} C^c.
\end{array}\eqno(3.4)
$$
Under the above dual BRST symmetry, the Lagrangian density transforms as:
$ \delta_{D} {\cal L_{B}} = \eta \partial_{\mu}
({\cal B}^a \partial^\mu \bar C^a)$. The on-shell nilpotent symmetry
transformations (3.2) lead to the Noether conserved current
$ J^\mu_{d} = [ F^{\mu\alpha a} + \eta^{\mu\alpha} (\partial \cdot A)^a ]
\varepsilon_{\alpha \rho} \partial^{\rho} \bar C^a $  which ultimately
leads to the dual BRST charge $ Q_{d} = \int dx
[ E^a \dot {\bar C^a} - (\partial \cdot A)^a \partial_{1} \bar C^a ] $.
Now using the partial integration and 
the equation of motion 
$ D_{0} E^a + \partial_{1} (\partial \cdot A)^a + i g f^{abc} C^b
\partial_{1} \bar C^c = 0$, this charge can be expressed as:

$$
\begin{array}{lcl}
Q_{(d,D)} &=& {\displaystyle \int}\; dx \bigl [\;E^a \dot {\bar C}^a 
- D_{0} E^a \bar C^a - i g f^{abc} \bar C^a \partial_{1} \bar C^b C^c
\;\bigr ],
\nonumber\\
&\equiv& {\displaystyle \int }\; dx \bigl [ \;{\cal B}^a \dot {\bar C}^a
- D_{0} {\cal B}^a \bar C^a - i g f^{abc} \bar C^a 
\partial_{1} \bar C^b C^c
\;\bigr ],
\end{array}\eqno(3.5)
$$
where the latter expression (for  $Q_{D}$) has been obtained
due to the validity of the equation of motion $ E^a = {\cal B}^a $. 
It can be checked that under the following off-shell nilpotent
($\delta_{AD}^2 = 0$) anti-dual BRST ($\delta_{AD}$) transformations
$$
\begin{array}{lcl}
&&\delta_{AD} A^a_{\mu} = - \eta \varepsilon_{\mu\nu} 
\partial^{\nu} C^a,
\quad \delta_{AD} C^a = 0, \quad \delta_{AD} \bar C^a 
= + i \eta {\cal B}^a,
\quad \delta_{AD} {\cal B}^a = 0, \nonumber\\
&&\delta_{AD} (\partial \cdot A)^a = 0, \;\quad 
\delta_{AD} B^a = 0,\quad
\;\delta_{AD} \bar B^a = 0, \;\quad 
\delta_{AD} E^a = \eta D_{\mu}
\partial^{\mu} C^a, 
\end{array}\eqno(3.6)
$$
the analogue of the Lagrangian density (2.7b) in 2D:
$$
\begin{array}{lcl}
{\cal L}_{\bar {\cal B}} = {\cal B}^a E^a 
- \frac{1}{2} {\cal B}^a {\cal B}^a
- \bar B^a (\partial \cdot A)^a + \frac{1}{2} 
(B^a B^a + \bar B^a \bar B^a)
- i D_{\mu}\bar C^a \partial^{\mu} C^a,
\end{array} \eqno(3.7)
$$
remains quasi-invariant because $\delta_{AD} {\cal L}_{\bar 
{\cal B}} = \eta \partial_{\mu} [ {\cal B}^a \partial^{\mu} C^a ] $. 
Using the Noether theorem, we obtain the conserved current 
as: $ J^{\mu}_{AD} = F^{\mu \alpha a} \varepsilon_{\alpha \beta}\;
\partial^{\beta} C^a + \varepsilon_{\alpha \beta} \partial^\beta C^a
\eta^{\mu\alpha} \bar B^a $ which leads to the anti-dual BRST charge
$Q_{AD} =  \int dx\; [\; {\cal B}^a \dot C^a - \bar B^a
\partial_{1} C^a \;] $. Using the partial
integration and exploiting the equations of motion
: $ \partial_{1} \bar B^a = i g f^{abc} \bar C^b \partial_{1} C^c - 
D_{0} {\cal B}^a $, we obtain
$$
\begin{array}{lcl}
Q_{AD} 
= {\displaystyle \int }\; dx \bigl [ \;{\cal B}^a \dot  C^a
- D_{0} {\cal B}^a  C^a + i g f^{abc}  C^a 
\partial_{1}  C^b \bar C^c
\;\bigr ].
\end{array}\eqno(3.8)
$$
It is straightforward to check that the above nilpotent and conserved
charges $Q_{r} ( r = B, AB, D, AD)$ are the generators of the transformations
(2.2), (2.9), (3.4) and (3.6) because these transformations can be concisely
expressed as
$ \delta_{r} \phi = - i \eta
[ \phi, Q_{r} ]_{\pm} $ where $(+)-$ stands for the (anti)commutator
corresponding to the generic field
$\phi$ being (fermionic)bosonic.\\

\noindent
{\bf 4 Symmetries generated by the Casimir operator}\\

\noindent
It is evident that the conserved and nilpotent charges $Q_{B}$ and
$Q_{D}$ are the fermionic symmetry generators corresponding to 
the transformations
$\delta_{B}$ and $\delta_{D}$ for the  Lagrangian density 
$ {\cal L_{B}}$ (cf. (3.3)). Their anticommutator $ W = \{ Q_{B}, Q_{D} \}$
will also generate a bosonic symmetry transformation $ \delta_{W}
= \{ \delta_{B}, \delta_{D} \} $. The following  transformations
$$
\begin{array}{lcl}
&&\delta_{W} \bar C^a = 0, \quad \delta_{W} C^a = 0, \quad
\delta_{W} B^a = 0, \quad \delta_{W} {\cal B}^a = 0, \nonumber\\
&& \delta_{W} A_{\mu}^a = \kappa \bigl [\; D_{\mu} {\cal B}^a
+ \varepsilon_{\mu\nu} \partial^\nu B^a - i g f^{abc} \varepsilon_{\mu\nu}
\partial^\nu \bar C^b C^c \;\bigr ], \nonumber\\
&& \delta_{W} E^a = - \kappa \bigl [ \;D_{\mu} \partial^{\mu} B^a
+  \varepsilon^{\mu\nu} D_{\mu} D_{\nu} {\cal B}^a - i g f^{abc}
D_{\mu} (\partial^{\mu} \bar C^b C^c) \;\bigr ], \nonumber\\
&& \delta_{W} (\partial \cdot A)^a = \kappa \bigl [\; \partial_{\mu} D^\mu
{\cal B}^a + i g f^{abc} \varepsilon^{\mu\nu} \partial_{\mu} 
\bar C^b \partial_{\nu} C^c \;\bigr ],
\end{array} \eqno(4.1)
$$
(with bosonic transformation parameters $ \kappa = - i \eta \eta^{\prime} $)
are indeed the symmetry transformations because the Lagrangian density
${\cal L_{B}}$ transforms to a total derivative as: 
$$
\begin{array}{lcl}
\delta_{W} {\cal L_{B}} = \kappa \partial_{\mu} \bigl [ B^a 
D^\mu {\cal B}^a
 - {\cal B}^a \partial^\mu B^a + i g f^{abc} ({\cal B}^a \partial^{\mu}
\bar C^b + \varepsilon^{\mu\nu} \partial_{\nu} \bar C^a B^b) C^c \bigr ].
\end{array}\eqno(4.2)
$$
In the expression for the bosonic transformation parameter $\kappa =
- i \eta \eta^{\prime}$, the fermionic parameters $\eta$ and $\eta^{\prime}$
correspond to the transformations generated by $Q_{B}$ and $Q_{D}$.
It will be noticed that, out of three basic fields, the ghost- and antighost
fields do not transform under $\delta_{W}$ and the gauge boson field 
$A_{\mu}^a$ transforms to its own equation of motion: $ D_{\mu} {\cal B}^a
+ \varepsilon_{\mu\nu} \partial^\nu B^a - i g f^{abc} \varepsilon_{\mu\nu}
\partial^\nu \bar C^b C^c (= 0)$. 
The generator of the above transformations
(4.1) is a conserved charge $W$ given by the following expression:
$$
\begin{array}{lcl}
W = \int \; dx \;\bigl [\; {\cal B}^a \dot B^a - B^a D_{0} {\cal B}^a
- i g f^{abc} ( {\cal B}^a \dot {\bar C^b} - \partial_{1} \bar C^a
B^b) C^c \;\bigr ].
\end{array}\eqno(4.4)
$$
Since in the BRST formalism, there are two more conserved and nilpotent 
charges $Q_{AB}$ and $Q_{AD}$, the anticommutator of these two can also
define operator $W$. Concentrating on the Lagrangian density (3.7), we can
check that the anticommutator $\{ \delta_{AB}, \delta_{AD} \}$ leads to the
following bosonic transformation $\delta_{W}$ (with bosonic transformation
parameter $ \kappa = - i \eta \eta^{\prime} $)
$$
\begin{array}{lcl}
&&\delta_{W} \bar C^a = 0, \quad \delta_{W} C^a = 0, \quad
\delta_{W} B^a = 0, \quad \delta_{W} {\cal B}^a = 0, \quad
\delta_{W} \bar B^a = 0,\nonumber\\
&& \delta_{W} A_{\mu}^a = \kappa \;\bigl [ \;- D_{\mu} {\cal B}^a
+ \varepsilon_{\mu\nu} \partial^\nu \bar B^a - i g f^{abc} 
\varepsilon_{\mu\nu}
\partial^\nu  C^b \bar C^c\; \bigr ], \nonumber\\
&& \delta_{W} E^a = - \kappa\; \bigl [ \;D_{\mu} \partial^{\mu} \bar B^a
- \varepsilon^{\mu\nu} D_{\mu} D_{\nu} {\cal B}^a
- i g f^{abc}
D_{\mu} (\partial^{\mu} C^b \bar C^c)\; \bigr ], \nonumber\\
&& \delta_{W} (\partial \cdot A)^a = -\kappa\; 
\bigl [\; \partial_{\mu} D^\mu
{\cal B}^a + i g f^{abc} \varepsilon^{\mu\nu} \partial_{\mu} 
\bar C^b \partial_{\nu} C^c \;\bigr ],
\end{array} \eqno(4.5)
$$
where $\eta$ and $\eta^{\prime}$ are the anticommuting transformation
parameters corresponding to $\delta_{AB}$ and $\delta_{AD}$ respectively.
It is straightforward to check that the Lagrangian density (3.7) undergoes
the following change under the transformations (4.5)
$$
\begin{array}{lcl}
\delta_{W} {\cal L}_{\bar B} = \kappa \;\partial_{\mu} \;\bigl [ 
 \bar B^a D^\mu {\cal B}^a
 - {\cal B}^a \partial^\mu \bar B^a + i g f^{abc} ({\cal B}^a \partial^{\mu}
 C^b + \bar B^a \varepsilon^{\mu\nu} \partial_{\nu}  C^b) \bar C^c\; \bigr ],
\end{array}\eqno(4.6)
$$
where use has been made of the identity
$$
\begin{array}{lcl}
i g f^{abc} {\cal B}^a D_{\mu} ( \partial^\mu C^b \bar C^c)
- i g f^{abc} \partial^{\mu} C^a D_{\mu} {\cal B}^b \bar C^c
= \partial_{\mu} [ i g f^{abc} {\cal B}^a \partial^\mu C^b \bar C^c ].
\end{array}
$$
The generator of the symmetry 
transformations (4.5) is
$$
\begin{array}{lcl}
W = \int \; dx \;\bigl [ \;{\cal B}^a \dot {\bar B^a} - \bar B^a 
D_{0} {\cal B}^a
+ i g f^{abc} ( {\cal B}^a \dot  C^b - \bar B^a \partial_{1}  C^b
) \bar C^c \;\bigr ].
\end{array}\eqno(4.7)
$$
Both the expressions for the conserved charge $W$ are equivalent because
they differ by a total space derivative when equations of motion for the
Lagrangian densities are used.

There are other simpler ways to obtain the expression for the 
analogue of the Laplacian operator $W$.
For instance, the symmetries (2.2), (2.9), (3.4) and (3.6) 
alone can be exploited
for the derivation of $W$. Since $Q_{r}$ (r = B, AB, D, AD) are the
generators of all these transformations, it can be seen that the following
transformations
$$
\begin{array}{lcl}
&&\delta_{B} Q_{D} \;= \;- i \eta \;\{ Q_{D}, Q_{B} \} \;\;=\;\; 
- i \eta W \; \;\;\equiv \;\delta_{D} Q_{B}, \nonumber\\
&& \delta_{AB} Q_{AD} = - i \eta \;\{ Q_{AD}, Q_{AB} \} = \;- i \eta W
\equiv \delta_{AD} Q_{AB},
\end{array}\eqno(4.8)
$$
also lead to the derivation of $W$. Furthermore,
these expressions for
$W$ can also be obtained from the anticommutators $\{ Q_{B}, Q_{D} \}$
or $\{ Q_{AD}, Q_{AB} \}$
by directly exploiting the basic canonical 
(anti)commutators for
 (3.3) and (3.7) which are juxtaposed as
$$
\begin{array}{lcl}
&& [ A^a_{0} (x,t), B^b (y,t) ] = i \delta^{ab} \delta (x - y),
\;\;\; \qquad
\;\;\;\;\;[ A^a_{0} (x,t), \bar B^b (y,t) ] = - i \delta^{ab} 
\delta (x - y),
\nonumber\\
&& [ A^a_{1} (x,t), {\cal B}^b (y,t) ] = i \delta^{ab} \delta (x - y), 
\;\;\;\;\qquad
\;\;\;\;[ A^a_{1} (x,t), {\cal B}^b (y,t) ]\; = i\; \delta^{ab}\;
 \delta (x - y),
\nonumber\\
&& \{ C^a (x,t), \dot {\bar C^b} (y,t) \} = \delta^{ab} \delta (x - y),
\;\;\;\qquad \;\;\;\;
\{ C^a (x,t), D_{0} \bar C^b (y,t) \} = \delta^{ab} \delta (x - y),
\nonumber\\
&&\{ \bar C^a (x,t), D_{0} C^b (y,t) \} = - \delta^{ab} \delta (x - y), 
\qquad
\{ \bar C^a (x,t), \dot C^b (y,t) \} = - \delta^{ab} \delta (x - y),
\end{array}\eqno(4.9)
$$
and all the rest of the (anti)commutators turn out to be zero. The above
canonical (anti)commutators lead to the derivation of the following extended
BRST algebra
$$
\begin{array}{lcl}
&& [ W, Q_{k}] = 0, k = B, D, AB, AD, g, \quad \nonumber\\
&&Q_{B}^2\; = \;Q_{AB}^2\; = \;Q_{D}^2\; = \;Q_{AD}^2\; = 0,\nonumber\\
&& \{ Q_{B}, Q_{D} \}\; = \;\{ Q_{AB}, Q_{AD} \}\; =\; W, \nonumber\\
&& i [ Q_{g}, Q_{B} ] = + Q_{B}, \;\;\;\quad\;\;
i [ Q_{g}, Q_{AB} ] = - Q_{AB}, \nonumber\\
&& i [ Q_{g}, Q_{D} ] = - Q_{D}, \qquad \;
i [ Q_{g}, Q_{AD} ] = + Q_{AD},
\end{array} \eqno(4.10)
$$
which is constituted by six conserved charges corresponding to six
symmetries present in the theory and all the rest of the (anti)commutators turn out to be zero. It is clear now that the operator
$W$ is the analogue of the Laplacian operator and is the Casimir
operator for the whole algebra. It can be also seen that the ghost number
for $Q_{B}$ and $Q_{AD}$ is $ + 1$ and that of $Q_{AB}$ and $Q_{D}$ is
$ - 1$. Now given a state $ |\psi>$ in the quantum Hilbert space with ghost 
number $n$ ( i.e., $ i Q_{g} | \psi> = n |\psi >$), it is straightforward 
to check that:
$$ 
\begin{array}{lcl}
i Q_{g}\; Q_{B} |\psi> &=& ( n + 1)\; Q_{B} |\psi>, \nonumber\\
i Q_{g}\; Q_{D} |\psi> &=& ( n - 1)\; Q_{D} |\psi>, \nonumber\\
i Q_{g}\; W \;    |\psi> &=& n\; W \;|\psi>.
\end{array}\eqno(4.11)  
$$
The above equation shows that the ghost number of the BRST exact state is
one higher (and that of the co-BRST exact state is one
lower) than the original state. This property is similar to the operation 
of an exterior derivative (and a dual exterior derivative) 
on a given differential form.
Thus, the geometrical quantities $d, \delta, \Delta $ find
their identifications in the language of symmetry properties that are
generated by $Q_{B}, Q_{D}$ and $W$.\\

\noindent
{\bf 5 Hodge decomposition theorem and topological invariants}\\

\noindent
A close look at the extended BRST algebra (4.10) and the considerations
of the ghost numbers for the BRST- and co-BRST exact states and harmonic
state (cf. (4.11)) allows one to implement the Hodge decomposition 
theorem in its full glory on any arbitrary state of the quantum Hilbert 
space (see, e.g., [4], [5], [13])
$$
\begin{array}{lcl}
|\psi >_{n} = |\omega >_{n} +\; Q_{B} |\; \theta >_{n-1} +\; Q_{D}\; 
| \chi >_{n+1}.
\end{array}\eqno(5.1)
$$
The above equation implies that  any arbitrary state $|\psi >_{n}$  with 
ghost number $n$ can be decomposed into a harmonic state $|\omega >_{n}$
($ W |\omega>_{n} = 0, Q_{B} |\omega>_{n} = 0, Q_{D} |\omega>_{n} = 0$),
a BRST exact state $ Q_{B} |\theta >_{n-1} $ and a dual-BRST exact state
$ Q_{D} | \chi >_{n +1} $.     
In fact, this equation is the analogue
of the mathematical statement on a compact manifold
that any arbitrary $p$-form $f_{p}$ can be
written as the sum of a harmonic form $\omega_{p}$ ($ \Delta \omega_{p} = 0,
d \omega_{p} = 0, \delta \omega_{p} = 0$), an exact form $ d g_{p-1}$ and
a co-exact form $\delta h_{p+1}$ due to the Hodge decomposition theorem:
$$
\begin{array}{lcl}
f_{p} = \omega_{p} + d g_{p-1} + \delta h_{p+1}.
\end{array}
$$   
Thus, the ghost number of a state in the quantum Hilbert space plays the
same role as the degree of a differential form defined on a compact 
manifold. It will
be noticed that the BRST cohomology can be defined either w.r.t. $Q_{B}$
or $Q_{D}$ or w.r.t. both.  To refine the BRST cohomology, however, we have
to choose a representative state from the
total states of (5.1) as a physical state. The harmonic states $|\omega>$
are very special for a given physical theory because they are finite in
number (see, e.g., [12]). Let us define our 
physical state as the harmonic state
(i.e., $ |phys> = |\omega>$). 
 By definition, such a state would satisfy
$$
\begin{array}{lcl}
 W |phys> = 0, \qquad Q_{B}\; | phys > = 0 , \qquad Q_{D}\; | phys > = 0.
\end{array} \eqno(5.2)
$$
In our earlier work on $U(1)$ gauge theory [17,18], it has been shown that
both the physical degrees of a single photon state in 2D can be gauged away
by the subsidiary conditions: $ Q_{B} |phys> = 0, Q_{D} |phys> = 0 $
alone. Thus, the operation of $W$ on this physical photon state 
becomes superfluous. In fact, the  Laplacian operator goes to zero
when the equations of motion are exploited and the theory becomes topological
in nature. In an analogous manner, it  turns out that both the expressions
for $W$ in (4.4) and (4.7) become

$$
\begin{array}{lcl}
W &=& {\displaystyle \int \; dx }\;
\frac{d} {dx} \;\bigl [ \; \frac{1}{2} \; {\cal B}^a {\cal B}^a
- \;\frac{1}{2}\; B^a B^a \; \bigr ]
\rightarrow  0 \;\;\;\;\quad \mbox{as} \;\;\;\;
x \rightarrow \pm \;\infty,\nonumber\\
&\equiv&
{\displaystyle \int \; dx }\;
\frac{d} {dx} \;\bigl [ \; \frac{1}{2} \; {\cal B}^a {\cal B}^a
- \;\frac{1}{2}\; \bar B^a \bar B^a \; \bigr ]
\rightarrow  0 \;\;\;\; \quad\mbox{as} \;\;\;\;
x \rightarrow \pm \;\infty,
\end{array}\eqno (5.3)
$$
as a consequence of the equation of motion for the gauge boson field
from (3.3)
$$
\begin{array}{lcl}
D_{0} {\cal B}^a - \partial_{1} B^a + i g f^{abc} 
\;\partial_{1} \bar C^b \;C^c &=& 0, \nonumber\\
D_{1} {\cal B}^a - \partial_{0} B^a 
+ i g f^{abc} \;\partial_{0} \bar C^b\; C^c &=& 0,
\end{array}\eqno(5.4a)
$$
and the same from Lagrangian density (3.7)
$$
\begin{array}{lcl}
D_{0} {\cal B}^a + \partial_{1} \bar B^a - i g f^{abc} 
 \bar C^b \partial_{1} C^c &=& 0, \nonumber\\
D_{1} {\cal B}^a + \partial_{0} B^a 
- i g f^{abc} \bar C^b \partial_{0} C^c &=& 0.
\end{array}\eqno(5.4b)
$$
It will be noticed that both the expressions for $W$ in (5.3)
are equivalent because $ B^a \partial_{1} B^a = \bar B^a \partial_{1}
\bar B^a - \partial_{1} X$ where $ X = i g f^{abc} \bar B^a C^b \bar C^c
+ \frac{1}{2} g^2 f^{abc} f^{amn} C^b \bar C^c C^m \bar C^n $. The vanishing
of the operator $W$ in (5.3) is the reflection of the fact that there are
no physical degrees of freedom left in the theory (as the BRST and co-BRST 
symmetries gauge away both the degrees of freedom of gauge boson in 2D) 
and it becomes topological in nature [11]. This situation should be 
contrasted with the interacting $U(1)$ gauge theory where Dirac fermions
couple to the gauge field. As it turns out, the Laplacian operator does
not go to zero even on the on-shell [22] because of the presence of
the fermionic degrees of freedom in the theory.

The topological nature of the theory is confirmed by the presence of
the topological invariants on the 2D manifold. The two sets of these 
invariants w.r.t. both the conserved and nilpotent charges $Q_{B}$
and $Q_{D}$ are (see, e.g., [10],[11],[23])

$$
\begin{array}{lcl}
I_{k} [C_{k}] = {\displaystyle \oint}_{C_{k}} \; V_{k}, \qquad
J_{k} [C_{k}] = {\displaystyle \oint}_{C_{k}} \; W_{k}, \qquad
k = 0, 1, 2
\end{array}\eqno (5.5)
$$
where $C_{k}$ are the k-dimensional homology cycles 
and the k-forms $V_{k}$ and $W_{k}$ are
$$
\begin{array}{lcl}
&&V_{0} = B^a\;C^a - \frac{i g}{2} 
f^{abc} \bar C^a C^b C^c, \;\;\;\qquad \;\;\;\;\;
W_{0} = {\cal B}^a\; \bar C^a, \nonumber\\
&&V_{1} = \bigl [ B^a A^a_{\mu} + i C^a D_{\mu} \bar C^a
\bigr ]\; dx^{\mu}, \;\;\;\qquad\;\;\;\;
W_{1} = \bigl [ \bar C^a \varepsilon_{\mu\rho} \partial^{\rho} C^a
- i\; {\cal B}^a A^a_{\mu} \bigr ] \;dx^{\mu}, \nonumber\\
&&V_{2} = i \bigl [ A^a_{\mu} D_{\nu} \bar C^a - \bar C^a
D_{\mu} A^a_{\nu} \bigr ] dx^{\mu} \wedge dx^{\nu},\; 
W_{2} = i \bigl [ \varepsilon_{\mu\rho} \partial^{\rho} C^a A^a_{\nu}
+ \frac{C^a}{2} \varepsilon_{\mu\nu} (\partial \cdot A)^a \bigr ]
dx^\mu \wedge dx^\nu.
\end{array}\eqno (5.6)
$$
It can be seen that $V_{0}$ and $W_{0}$ are BRST and co-BRST invariant
respectively and $V_{2}$ and $W_{2}$ are closed and co-closed respectively.
These invariants for ($ k = 1, 2 $) obey [23], [10]
$$
\begin{array}{lcl}
\delta_{B} V_{k} &=& \eta\; d\; V_{k-1}, \qquad\;\;
d = dx^{\mu}\;\partial_{\mu}, \nonumber\\
\delta_{D} W_{k} &=&  \eta\; \delta \; W_{k-1}, \;\qquad
\delta = i \;dx^{\mu}\; \varepsilon_{\mu\nu}\;
\partial^{\nu},
\end{array}\eqno (5.7)
$$
where $d$ and $\delta$ are the exterior and dual-exterior derivatives on
the 2D compact manifold. Unlike the $U(1)$ gauge theory [18] here there are
no specific transformations that can relate $I_{k}$ and $J_{k}$. Using
the on-shell nilpotent transformations (2.3) and (3.2), it is
interesting to verify that, modulo some total derivatives, the Lagrangian
density (3.1) can be written as the sum of BRST- and co-BRST invariant
parts:
$$
\begin{array}{lcl}
\eta {\cal L}_{b} = \frac{1}{2}\;\delta_{d}\; \bigl [ \;i E^a C^a \;\bigr ]
- \frac{1}{2}\;\delta_{b}\; \bigl [\; i (\partial \cdot A)^a \bar C^a\;
\bigr ].
\end{array}\eqno(5.8)
$$
Using the fact that conserved and nilpotent charges $Q_{r}$ (r = b, d) are 
the generator of transformations $ \delta_{r} \phi = - i \eta 
[ \phi, Q_{r} ]_{\pm},$ where $(+)-$ stands for the (anti)commutator
corresponding to the generic field $\phi$ being (fermionic)bosonic, it
can be seen that the Lagrangian density (5.8) can be recast as:
$ {\cal L}_{b} = \{ Q_{d}, S_{1} \} + \{ Q_{b}, S_{2} \} $ for
$ S_{1} = \frac{1}{2} E^a C^a $ and $ S_{2} = - \frac{1}{2} (\partial
\cdot A)^a \bar C^a$. This demonstrates that the topological theory
under consideration is similar in outlook as the Witten type theories [10].
There is a bit of difference, however. This is because of the fact
that there are two conserved and nilpotent charges in our discussion
whereas there exists only one conserved and nilpotent BRST charge in 
Ref. [10]. It is straightforward to check that the partition
functions and expectation values of the BRST invariants, co-BRST
invariants and the topological invariants are metric independent
\footnote{It will be noticed that we have taken here the flat Minkowski
metric. However, our arguments and discussions are valid for any
nontrivial metric. The metric independence of the path integral measure
for the topological field theories has been shown in Ref. [23].}. 
The main argument to show this fact in the framework of BRST 
cohomology is the requirement that $ Q_{b} |phys> = 0, Q_{d} |phys> = 0 $
(see, e.g., Ref. [11]) and the metric independence of the path integral
measure (see, e.g., Ref. [23]).\\

\noindent
{\bf 6 Conclusions}\\

\noindent
We have shown that the nilpotent symmetry transformations
under which the gauge-fixing
term remains invariant ($\delta_{D} [(\partial \cdot A)^a (\partial
\cdot A)^a] = 0 $) is the dual BRST symmetry in contrast to the usual
BRST symmetry under which the total kinetic energy term remains
invariant ($\delta_{B} [F^{\mu\nu a} F^a_{\mu\nu}] = 0$). The 
anticommutator of these two symmetries corresponds to a symmetry
(generated by the Laplacian(Casimir) operator) under which the
ghost fields do not transform ($\delta_{W} C^a = \delta_{W} \bar C^a
= 0 $) and the gauge connection transforms to its own equation of
motion. Thus, this symmetry becomes trivial on the on-shell. This 
triviality is connected with the topological nature of the 2D free 
non-Abelian
gauge theory as the BRST- and co-BRST symmetries are good enough to
gauge away both the degrees of freedom of the gauge boson.
In fact, the Laplacian operator goes to zero when the
equations of motion are exploited. The
on-shell expression of the Laplacian operator encompasses the degrees
of freedom left in the theory. In the case of $U(1)$ gauge field
coupled to the Dirac fermion, it has been shown that the Laplacian
operator does not go to zero even on the on-shell and its expression 
contains only the fermionic degrees of freedom [22]. Furthermore, it 
has been demonstrated that the co-BRST transformation on the $U(1)$ 
gauge field
corresponds to the quantum chiral transformation on the Dirac fermions
in 2D. This symmetry, therefore, might shed light on the ABJ anomaly
in 2D and might provide clue to the consistency of the ``anomalous'' 
gauge theory in 2D (see, e.g., [24,25]). It will be interesting to 
study the BRST
cohomolgy when non-Abelian gauge field is coupled to the matter fields
and generalise these understandings to the case of
gauge theories in physical four dimensions.\\

\baselineskip = 12pt

\noindent {\bf References}
\begin{enumerate} 
\item  P. A. M. Dirac, Lectures on Quantum Mechanics,
       (Yeshiva University Press, New York, 1964).
\item  K. Sundermeyer, 1982  Constrained Dynamics, Lecture Notes
       in Physics, Vol. 169, (Springer-Verlag, Berlin, New York, 1982).
\item  K. Nishijima, in: Progress in Quantum Field Theory, eds.
         Ezawa
       H. and Kamefuchi S., ( North- Holland, Amsterdam, 1986 ), p. 99.
\item  M. Henneaux  and C. Teitelboim ,  
       Quantization of Gauge Systems, 
       (Princeton University Press, Princeton, 1992).
\item  N. Nakanishi and I. Ojima, Covariant Operator Formalism of
       Gauge Theories and Quantum Gravity, 
       (World Scientific, Singapore, 1990).
\item  J. Gomis, J. Paris and S. Samuel, Phys. Rep. 259 (1995) 1.
\item  E. Witten, Mod. Phys. Lett. A5  (1990) 487.
\item  I. A. Batalin and I. V. Tyutin, Int. J. Mod. Phys. A6 (1991) 3255,
       I. A. Batalin, S. L. Lyakhovich and I. V. Tyutin, Mod. Phys. Lett.
       A7 (1992) 1931, Int. J. Mod. Phys. A10 (1995) 1917.
\item  A. S. Schwarz, Lett. Math. Phys. 2 (1978) 217.
\item  E. Witten, Commun. Math. Phys. 121 (1989) 351.
\item  D. Birmingham, M. Blau, M. Rakowski and G. Thompson, Phys. Rep.
       209 (1991) 129.
\item  S. Mukhi  and N. Mukunda,  Introduction  to Topology,
       Differential Geometry and Group Theory for Physicists, (Wiley Eastern
       Ltd., New Delhi, 1990).
\item  J. W. van Holten,  Phys. Rev. Lett.  64 (1990) 2863;
       Nucl. Phys. B339 (1990) 158.
\item  T. Eguchi, P. B. Gilkey and A. J. Hanson,  
       Phys. Rep.  66 (1980) 213. 
\item  D. McMullan  and M. Lavelle,  Phys. Rev. Lett.  71 (1993) 3758,
       {\it ibid.}  75 (1995) 4151,
       V. O. Rivelles, Phys. Rev. Lett.  75 (1995) 4150; 
       Phys. Rev. D 53 (1996) 3257,
       H. S. Yang  and B. -H. Lee,  J. Math. Phys.
       37 (1996) 6106,
       R. Marnelius,  Nucl. Phys.  B494 (1997) 346, H. Aratyn, J. Math. 
       Phys. 31 (1990) 1240.
\item  T. Zhong and D. Finkelstein,   Phys. Rev. Lett.
       73 (1994) 3055;   {\it ibid.}  75 (1995) 4152.
\item  R. P. Malik, On the BRST cohomology in $U(1)$ gauge theory:
       hep-th/9808040.
\item  R. P. Malik, Topological aspects in $U(1)$ gauge theory:
       hep-th/9902146.
\item  S. Weinberg,  
       The Quantum Theory of Fields: Modern Applications
       V.2 (Cambridge University Press, Cambridge, 1996).
\item  G. Curci and R. Ferrari, Phys. Lett. 63B (1976) 51;
       Nuovo Cimento 32A (1976) 151, L. Bonora and M. Tonin, Phys.
       Lett. 98B (1981) 48.
\item  C. Chryssomalakos, J. A. de Azcarraga, A. J. Macfarlane and J. C.
       Perez Bueno, Higher order BRST and anti-BRST operators and
       cohomology for compact Lie algebras: hep-th/9810212.
\item  R. P. Malik,  Dual BRST Symmetry in QED: 
       hep-th/ 9711056.
\item  R. K. Kaul and R. Rajaraman, Phys. Lett. 265B (1991) 335,
       {\it ibid.} 249B (1990) 433.   
\item  R. Jackiw  and R. Rajaraman,  Phys. Rev. Lett.  
       54 (1985) 1219.
\item  R. P. Malik,  Phys. Lett. 212B (1988) 445.
\end{enumerate}
\end{document}